\documentclass[12pt]{article}

\usepackage[T1]{fontenc}
\usepackage{lmodern}
\usepackage[utf8]{inputenc}
\usepackage[american]{babel}

\usepackage{geometry}
\usepackage{setspace}
\usepackage{microtype}
\usepackage[section]{placeins}

\usepackage{amsmath, amssymb}
\usepackage{graphicx}
\usepackage{booktabs}
\usepackage{threeparttable}
\usepackage{array}
\usepackage{tabularx}
\usepackage{longtable}
\usepackage{caption}
\usepackage{float}
\usepackage{ragged2e}
\usepackage{siunitx}
\usepackage{makecell}

\graphicspath{{figures/}{./}}
\setkeys{Gin}{width=\linewidth,keepaspectratio}
\newcommand{\safeincludegraphics}[2][]{\includegraphics[#1]{#2}}

\usepackage{xcolor}
\definecolor{caseRed}{HTML}{b2182b}
\definecolor{ctrlBlue}{HTML}{2166ac}
\definecolor{hubSalmon}{HTML}{f4a582}
\definecolor{giverPink}{HTML}{e377c2}
\definecolor{recvrCyan}{HTML}{17becf}

\newcolumntype{Y}{>{\RaggedRight\arraybackslash}X}
\renewcommand{\arraystretch}{1.15}

\usepackage{authblk}


\usepackage{csquotes}
\usepackage[style=apa, backend=biber, maxcitenames=2, uniquelist=false]{biblatex}
\DeclareLanguageMapping{american}{american-apa}

\usepackage{xurl} 
\usepackage[unicode,breaklinks=true]{hyperref}
\hypersetup{
  colorlinks=true,
  urlcolor=blue,
  linkcolor=black,
  citecolor=black
}

\title{Citation Cliques in Low Impact Journals}
\author{Panagiotis-Alexios Spanakis}
\author{Grigorios Alexandrou}
\author{Diomidis Spinellis}
\affil{Department of Management Science and Technology, 
Athens University of Economics and Business (AUEB), Athens, Greece}
\date{}

\begin{document}
\maketitle

\begin{abstract}
  This exploratory study examines how low-impact journals, defined through subject-normalized Eigenfactor percentiles, are associated with denser and more reciprocating patterns of author-to-author citations.
  Using Crossref records, we assign journals to broad subject areas, compute subject-specific Eigenfactor scores, propagate venue quality to works and authors, match authors in low-- (Case) versus high-influence (Control) venues by subject and $h_5$, and analyze citation edges for cohesion and anomalies.
  Across a 10\% sample of \num{9431} matched pairs, authors in low-impact venues exhibit significantly higher cohesion: 6.7$\times$ higher co-author citation rates and 4.7$\times$ higher reciprocity in the aggregate Case--Control comparison.
  A subject-aware hybrid detection pipeline flags 277 outliers with 93.5\% Case purity; these outliers display an 11$\times$ clique-strength lift relative to non-outliers, revealing a stark ``Two Worlds'' segregation ($r=0.71$) where low-impact venues operate as closed citation economies.
  The largest detected component ($n=23$) displays a hub-and-spoke topology in which peripheral ``Sycophants'' funnel citations to central ``Beneficiaries'' through coordinated bursts, confirming a directed flow imbalance rather than reciprocal exchange among equals.
  Overall, cohesion, rather than broad asymmetry, accounts for the main Case--Control differences, suggesting that low-impact venues foster segregated, inward-looking citation economies that distort bibliometric indicators.
\end{abstract}


\section{Introduction}
\label{sec:intro}
Quantitative indicators such as citation counts, the $h$-index, and journal metrics have become central to the assessment of scholarly impact \parencite{Gar06, Hirsch2005, Hicks2015}. These measures offer convenient proxies for influence, but they also create incentives for strategic behavior and manipulation \parencite{Goo84,Biagioli2020,Baccini2019}. When deployed uncritically, citation-based indicators risk rewarding volume or visibility over substance, with unintended consequences for research integrity and evaluation systems.

Beyond well-known practices such as excessive self-citation \parencite{VanNoorden2019,Wilhite2012}, more coordinated behaviors have emerged. Authors may form tightly connected \emph{citation cliques} --- groups that disproportionately reference one another, often within low-impact journals as defined by metrics such as the Journal Impact Factor~\parencite{Garfield1972} or subject-normalized Eigenfactor percentiles \parencite{West2008}. Such practices distort indicators, undermine trust in the scholarly record, and complicate fair comparison across fields. These dynamics exemplify what has recently been described as \emph{citation orchestration} \parencite{Evdaimon2024}, where citations serve strategic rather than intellectual functions.

Low-impact journals provide fertile ground for these dynamics. Their emphasis on publication volume, rapid turnaround, and relaxed editorial oversight \parencite{Beall2012, Abalkina2025} creates environments where abnormal citation patterns can thrive. Systemic distortions such as national \emph{home bias} \parencite{Qiu2025} further interact with venue incentives, amplifying the risk that rankings and impact assessments reflect coordination rather than genuine uptake of ideas.

This exploratory study investigates whether low-impact journals are significantly associated with cliques and abnormal citation patterns. Using large-scale bibliographic data, we implement a subject-aware pipeline that links journal influence to authors,
matches researchers publishing in Bottom-- and Top-tier impact journals
of comparable productivity (by $h_5$ index), and analyzes their citation networks for cohesion and anomalies.

We find that extreme anomalies cluster predominantly among authors publishing mostly in low-impact venues: they concentrate in unusually dense, inward-looking structures and form compact outlier groups that set them apart from their Control counterparts.
These results provide empirical evidence that venue incentives can shape citation behavior in ways that distort bibliometric indicators.
Throughout this paper, authors publishing primarily in low-impact venues are referred to as \emph{Cases}, and their matched counterparts in high-impact venues as \emph{Controls}.
Terms such as ``syndicate,'' ``sycophant,'' and ``beneficiary'' are used as \emph{structural labels} describing observed network roles and do not imply intent or misconduct.
Similarly we use the terms Control and Case to classify authors publishing
mainly in high-- or low-impact journals, again without implying any
judgement or view regarding the authors' and their works' scientific worth.

\section{Related Work}
\label{sec:relwork}

Prior research has highlighted the fragility of citation-based indicators when used as proxies for scholarly quality.
Performance measures not only shape incentives~\parencite{Goo84},
but also invite gaming, from excessive self-citation to coordinated cartels \parencite{Biagioli2019,Biagioli2020}.
The wider literature on the “metrics culture” documents how quantitative indicators reshape scholarly communication,
creating feedback loops between evaluation systems and researcher behavior \parencite{Hicks2015, Biagioli2020}.
These mechanisms underpin the emergence of citation cartels~\parencite{Fister2016,Catanzaro2024},
inflated impact factors~\parencite{Ioannidis2024extreme,Catanzaro2024},
paper mills~\parencite{Christopher2021,Abalkina2025}, AI-generated papers~\parencite{Spi25d},
and other forms of metric manipulation that erode the interpretive value of bibliometric indicators.

Venue-level frameworks were developed to mitigate some of these distortions. The Eigenfactor metric \parencite{West2008} introduced a network-based approach that discounts journal self-citations and weights each citation by the prestige of the citing venue. This recursive weighting sought to capture influence rather than popularity and to counteract local citation loops. However, later analyses have shown that such corrections remain sensitive to database coverage, field size, and reference density \parencite{Baccini2019}, indicating that structural biases persist even in sophisticated venue-level measures.

At the author level, organized citation exchanges have been formalized as \emph{citation orchestration} \parencite{Evdaimon2024}. This framework quantifies concentrated citation flows through indicators such as the ratio of total citations to the squared author's $h$-index ($C/h^2$), the number of distinct authors accounting for half of one's citations ($A50\%C$), and the count of hyper-collaborators sharing over fifty joint papers ($A50$). Together, these metrics reveal unusually cohesive collaboration networks that may signal collective metric inflation. Earlier studies of citation cartels and excessive self-referencing \parencite{Fister2016,Baccini2019,Kojaku2021,Ioannidis2025precocious,heneberg2016} align with this view, emphasizing how reciprocal ties among small groups of authors can yield disproportionate bibliometric visibility. Such coordinated behaviors blur the line between collaboration and manipulation, challenging the assumption that citation networks represent organic intellectual influence.

Systemic and infrastructural distortions further complicate these dynamics. National \emph{home bias} inflates domestic citation counts, particularly in large scientific systems such as China's \parencite{Qiu2025}, while technical manipulation has emerged through practices such as ``sneaked references'', which are inserted into metadata but absent from article text \parencite{Besancon2024}. Industrial-scale paper mills produce fabricated papers that populate citation networks with artificial linkages \parencite{Abalkina2025}. Recent investigations reveal even more elaborate schemes, including the creation of fictitious author identities and peer reviewers to infiltrate legitimate journals. As reported by \textcite{Naddaf2025}, these fake personas --- sometimes maintained across multiple journals --- enable mills to generate both papers and favorable reviews, prompting publishers to explore researcher identity verification frameworks and persistent identifier checks such as ORCID trust markers.

The present research extends this discourse by systematically examining the relationship between author-level citation behaviors and the influence level of the publication venue, while controlling for individual productivity ($h_5$ index) and disciplinary norms through a matched Case--Control design. While previous studies have primarily investigated the orchestration of specific bibliometric indicators \parencite{Evdaimon2024} or semantic irregularities within article metadata \parencite{Liu2024}, this study identifies \emph{structural cohesion}, which manifests as dense local clustering and triadic closure, as the primary diagnostic signature of citation anomalies within low-impact journals. By implementing a subject-aware isolation forest on author-to-author networks across a 10\% sample of \num{9431} matched pairs, the analysis demonstrates that such cliques constitute structural byproducts of venue-level incentives. Consequently, this approach provides a network-centric diagnostic framework for identifying organized citation manipulation practices that evade traditional bibliometric oversight by focusing on network closure patterns rather than simpler citation calculations.

\section{Methodology}
\label{sec:methods}

\begin{figure}[H]
  \centering
  \safeincludegraphics{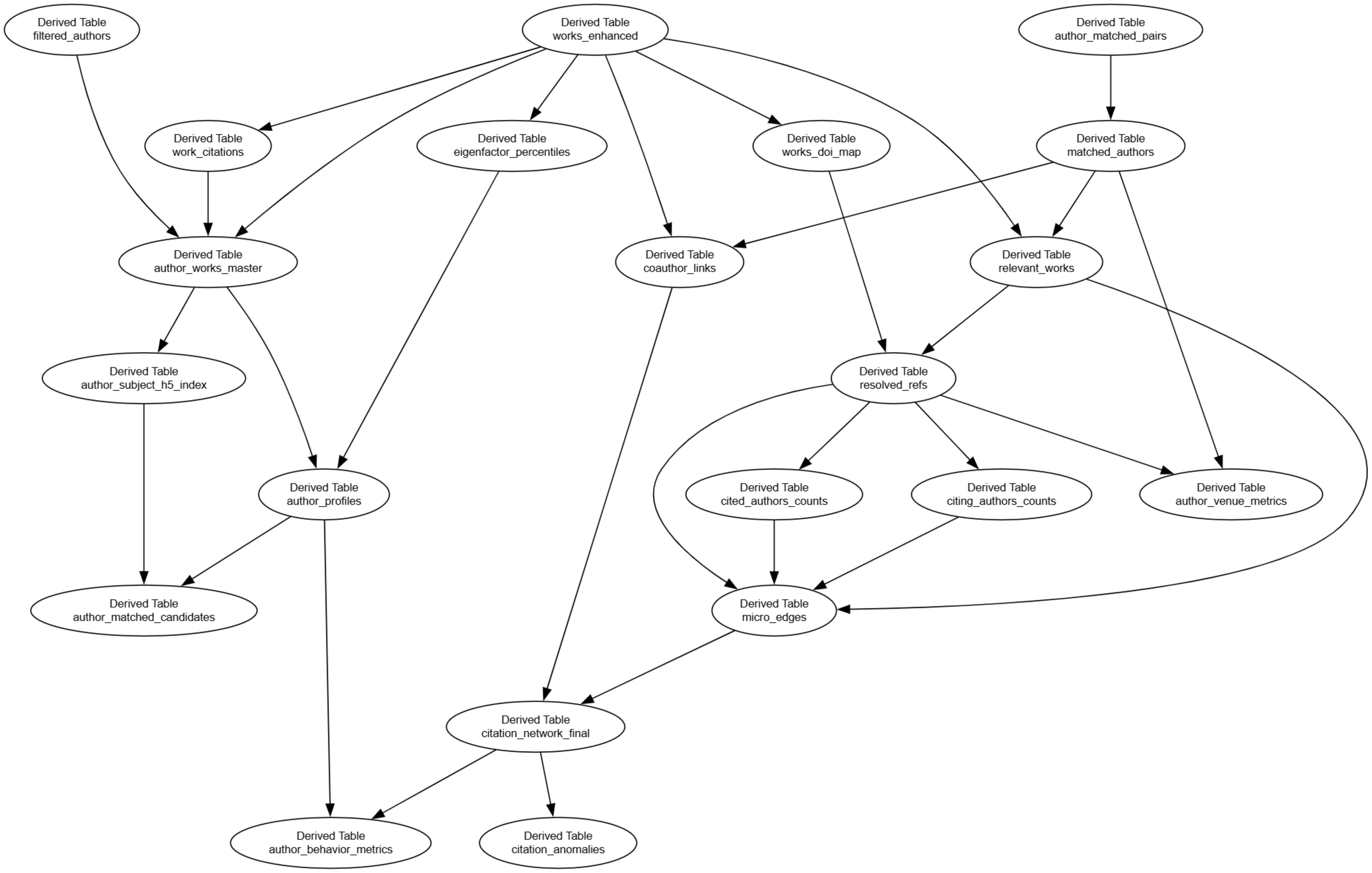}
  \caption{Analysis DAG for the social-network stage. Ellipses represent derived tables and arrows indicate dependencies. Curated inputs feed \texttt{citation\_network\_final} and downstream aggregates.}
  \label{fig:analysis-dag}
\end{figure}

We developed a subject-area-aware pipeline linking journal quality to author citation behavior. The main steps were: (i) journal subject classification, (ii) computation of subject-specific journal Eigenfactor scores, (iii) propagation of venue impact to works and authors, (iv) matched pairing of Case and Control authors based on journal impact and productivity, and (v) construction of author--author citation networks with cohesion and anomaly features. Figure~\ref{fig:analysis-dag} illustrates the main analysis subgraph of this pipeline.

To ensure consistency, replicability, and correctness
across all stages of the data pipeline before statistical analysis,
these were orchestrated through \emph{simple-rolap} \parencite{Spi24e}
and unit-tested through \emph{RDBUnit} \parencite{Spinellis2024RDBUnit}.

\subsection{Data and Preprocessing}

\textbf{Data collection.}
Crossref metadata were ingested via \textit{alexandria3k} \parencite{Spinellis2023}, covering works published between 2020 and 2024. We retained DOIs, years, titles, abstracts, references, author identifiers (ORCID), and normalized ISSNs for journals. Duplicates and inconsistencies were cleaned, and derived layers linked works to authors, journals, and citations.
Appendix~\ref{app:sampling} documents the full data derivation pipeline and the filtering decisions applied at each stage.

\textbf{Journal Subject Classification.}
Each journal was assigned to one of five broad subject areas:
1) Health Sciences,
2) Life Sciences,
3) Physical Sciences,
4) Social Sciences \& Humanities, and
5) Multidisciplinary,
using an LLM as an annotator \parencite{LLMGuidelines2024}.
This taxonomy mirrors the broad subject areas of the All Science Journal
Classification (ASJC) scheme used by Scopus \parencite{Burnham2006}, one of the
most widely adopted classification frameworks in scientometrics. Separating subject
areas is essential for computing meaningful within-field Eigenfactor percentiles,
as citation practices vary substantially across disciplines \parencite{West2008}.

Classification was performed using OpenAI \texttt{gpt-4.1}
(snapshot \texttt{gpt-4.1-2025-04-14}, accessed June~2025) with zero temperature and
few-shot exemplars. The model processed journals in batches of ten ISSNs per API
call, supplying for each journal up to five representative works ordered by citation
count, so that the most influential works appeared first. Prompts and decision rules
prioritized abstracts over titles and journal names; when an abstract was
unavailable, classification relied on the journal name and title alone.
To ensure reproducibility, we used a fixed random number generation seed
and cached predictions to JSON files.
The full prompt template and representative input--output pairs are
included in the replication package \parencite{dataset_zenodo}.
No open-weight baseline model was evaluated; this is acknowledged
as a limitation (Section~\ref{sec:limitations}).

\textbf{Eigenfactor Computation.}
To quantify journal impact, we computed subject-specific Eigenfactor scores \parencite{West2008} on journal citation networks (2020--2024), excluding self-citations. The computation used standard power iteration and produced normalized influence scores within each subject.

\textbf{Defining Venue Impact.}
Within each subject, journals in the top quartile of Eigenfactor were labeled \emph{high-impact}, and those in the bottom quartile \emph{low-impact}. Labels propagate to works (by ISSN) and to authors through their portfolios. Alternative thresholds and persistence checks were evaluated for robustness.

\subsection{Author Matching and Network Construction}

\textbf{Author Tiering and Matching.}
Authors were classified as \textbf{Control} if most of their output was in high-impact journals, \textbf{Case} if concentrated in low-impact journals, and \textbf{Mixed} otherwise.
Mixed-tier authors were excluded from the matched-pair analysis to ensure clean separation between venue quality levels.
We then matched Case authors to Control authors of similar productivity (subject-specific $h_5$ index), ensuring comparability in output volume and field.

\textbf{Sample and Availability.}
The final analytical dataset comprises a 10\% sample of \num{9431} matched Case--Control author pairs
(see Table~\ref{tab:sampling-funnel} for the per-subject breakdown).
All pair-level and network-level aggregates were computed per subject area and combined
only after validation of within-subject consistency.
All processed data tables are available through the public replication repository
described in Section~\ref{sec:data}.

\textbf{Citation Network and Metrics.}
For matched authors, we constructed directed author-to-author citation networks (2020--2024).
Each citation edge between a citing work with $n_c$~authors and a cited work with $n_a$~authors received weight $w = 1/(n_c \times n_a)$, correcting for team size.
From these edges we derived cohesion metrics (self-citation, co-author citation, clustering, triangles, $k$-core, eigenvector centrality) and pairwise anomaly metrics (reciprocity, asymmetry, velocity, citation bursts). These measures capture both dense reciprocal structures and asymmetric one-way flows.
During exploratory analysis, raw pairwise metrics (maximum asymmetry, average velocity, and maximum burst) were also computed; in the formal pipeline they were replaced by normalised variants that are comparable across activity levels: \emph{citation balance} captures net giver/receiver bias as a bounded $(-1,+1)$ ratio, while \emph{normalised burst intensity} adjusts the largest single-peer citation weight by total incoming volume.
Table~\ref{tab:features} (Appendix~\ref{app:features}) lists the 14 features employed for statistical testing and modelling.

\subsection{Statistical Analysis and Outlier Detection}

\textbf{Statistical Analysis and Robustness.}
Matched pairs were compared using Wilcoxon signed-rank tests \parencite{Wilcoxon1945}, with Benjamini-Hochberg FDR correction \parencite{Benjamini1995}.
Effect sizes were quantified via Cliff's $\delta$ \parencite{Cliff1993} and Cohen's $d$ \parencite{Cohen1988},
with confidence intervals estimated through bootstrap resampling \parencite{Efron1993}.
Robustness checks included permutation tests, placebo pairing, random-effects meta-analysis across subjects,
and alternative outlier detectors (Isolation Forest \parencite{Liu2008}, Local Outlier Factor \parencite{Breunig2000}, One-Class SVM).
Sensitivity analysis utilized z-score thresholds ($1\sigma, 2\sigma, 3\sigma, 4\sigma$) relative to the control baseline to establish interpretable detection cutoffs.
Community detection used the Louvain method \parencite{Blondel2008}, which optimises the modularity criterion \parencite{NewmanGirvan2004}.

\textbf{Outlier Detection Procedure.}\label{sec:outlier-proc}
We developed a subject-aware hybrid outlier detection strategy linking multivariate anomalies to specific cohesion signatures.
The pipeline combines an Isolation Forest (IF) \parencite{Liu2008} with a domain-specific Cohesion Composite Score.
The IF model used standard hyperparameters (\texttt{n\_estimators=200}, \texttt{contamination=0.01}) to capture general anomalies.
The Cohesion Score was computed as the standardized sum of the four most discriminative features: co-author citation rate, clique strength, reciprocity, and outgoing HHI\footnote{The Herfindahl-Hirschman Index (HHI) measures concentration. In this context, it quantifies the extent to which an author repeatedly cites the same narrow set of recipients, indicating a lack of source diversity.}.

We computed 14 behavioural features per author (Table~\ref{tab:features}).
Of these, normalised burst intensity was available for only \num{7711} of the \num{9431} matched pairs (18\% missing) because many authors recorded no burst events in the study window.
Statistical tests handle this by dropping incomplete pairs per metric; however, the Isolation Forest requires a complete feature matrix, and zero-imputing 18\% of values would inject a spurious signal.
The model was therefore trained on the remaining 13 features.
All features were standardised (zero mean, unit variance) and multiplied by hand-tuned weights reflecting Random-Forest importance and Wilcoxon effect sizes (e.g., Co-author Rate weighted 4.0); authority metrics were sign-inverted so that lower values indicate greater suspicion.
An author was flagged as an outlier only if they satisfied both conditions: an IF anomaly score above the threshold \textbf{AND} a Cohesion z-score $> 4\sigma$ (relative to the Control baseline).
Sensitivity tests evaluated thresholds from $1\sigma$ to $4\sigma$ to confirm the robustness of detected syndicates.
Outlier labels were merged back into the unified author-feature table for downstream network analyses (results in Section~\ref{sec:outlier-results}).

\section{Findings}
\label{sec:findings}

This section presents the main results of the matched Case--Control analysis.
We first examine overall and field-specific patterns of cohesion and asymmetry,
then derive unsupervised behavioral archetypes and identify statistical outliers
using a subject-aware Isolation Forest.
Together, these analyses show that cohesion-rather than sustained one-sided citation flows
is the primary structural distinction between authors publishing in low-- and high-influence venues.
The main network features used were
the \emph{clustering coefficient} (density of local interconnections);
\emph{triangles} (closed three-way citation structures);
the \emph{k-core number} (depth of embedding in dense subnetworks);
\emph{betweenness centrality} (how often an author lies on shortest paths, a proxy for brokerage/coordination potential);
\emph{velocity} (rate of citation accumulation);
\emph{burst} (sudden within-window surges of citations); and
\emph{asymmetry} (imbalance in giving vs.\ receiving citations).

\subsection{All Subjects Combined}
\label{sec:overall}

\begin{figure}[H]
  \centering
  \safeincludegraphics[width=\linewidth]{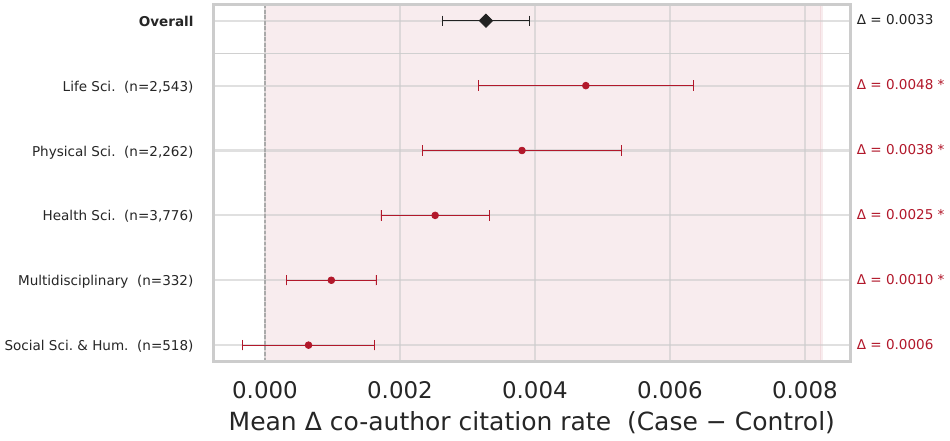}
  \caption{\textbf{Co-author citation gap by subject field.}
    Forest plot of mean paired difference
    $\Delta = \bar{r}_{\text{Case}} - \bar{r}_{\text{Control}}$
    in co-author citation rate, with 95\% confidence intervals.
    Marker $\bullet$ = field estimate; $\blacklozenge$ = overall.
    $\Delta$ labels report rounded effect sizes.
    The dashed vertical line marks zero (no difference);
    the pale red shading covers the positive region where
    Case authors exceed Controls.}
  \label{fig:forest_plot}
\end{figure}

At the aggregate level, authors publishing primarily in low-impact venues display markedly stronger \emph{cohesion}:
on average, they cite co-authors \textbf{6.7$\times$ more frequently} ($p=9.1 \times 10^{-66}$), exhibit \textbf{4.7$\times$ higher reciprocity rates} ($p=2.7 \times 10^{-44}$), and show \textbf{3.1$\times$ greater outgoing concentration} (tunnel vision, $p \approx 0$).
Figure~\ref{fig:forest_plot} displays the mean paired difference in co-author citation rates across subject fields.

Table~\ref{tab:metrics-overview} summarises the aggregate Case--Control
contrast across all \num{9431} matched pairs and illustrates it with a
representative pair drawn from Health Sciences.

\begin{table}[H]
  \centering
  \caption{\textbf{Case--Control metric averages and example pair.}
    Mean values computed over all \num{9431} matched pairs.
    The example pair is drawn from Health Sciences (Subject~1),
    both non-outlier.}
  \label{tab:metrics-overview}
  \small
  \setlength{\tabcolsep}{4pt}
  \begin{tabular}{@{} l S[table-format=1.4] S[table-format=1.4]
                      S[table-format=1.4] S[table-format=1.4] @{}}
    \toprule
    & \multicolumn{2}{c}{\textbf{Mean}} & \multicolumn{2}{c}{\textbf{Example pair}} \\
    \cmidrule(lr){2-3} \cmidrule(l){4-5}
    \textbf{Metric} & {Case} & {Control} & {Case} & {Control} \\
    \midrule
    ORCID
      & {---} & {---}
      & \multicolumn{1}{c}{\texttt{\dots3810}}
      & \multicolumn{1}{c}{\texttt{\dots8680}} \\
    \addlinespace
    Co-author cit.\ rate  & 0.0040 & 0.0006 & 0.0031 & 0.0005 \\
    Self-citation rate    & 0.0278 & 0.0107 & 0.0272 & 0.0104 \\
    Reciprocity rate      & 0.0042 & 0.0009 & 0.0290 & 0.0256 \\
    Outgoing HHI          & 0.0868 & 0.0277 & 0.0168 & 0.0256 \\
    Clustering coeff.     & 0.0612 & 0.0525 & 0.3000 & 0.0000 \\
    Clique strength       & 0.0011 & 0.0001 & 0.0009 & 0.0000 \\
    Citation balance      & 0.7580 & 0.9642 & 0.9638 & 1.0000 \\
    Eigenvector centr.\ ($\times 10^{5}$) & 1.18 & 15.36 & 0.00 & 0.00 \\
    $k$-core number       & 0.7080 & 1.2170 & 3.0000 & 1.0000 \\
    Citation entropy      & 0.1210 & 0.2640 & 1.0990 & 0.0000 \\
    Journal endogamy      & 0.0727 & 0.0887 & 0.0886 & 0.0984 \\
    \bottomrule
  \end{tabular}
\end{table}

The cohesion gap is not driven by a few extreme outliers but by a broad rightward shift across all four dimensions (co-author citation rate, self-citation rate, outgoing HHI, and local clustering).
Permutation testing (\num{10000} label shuffles of Case--Control assignments) independently confirms the effect is far outside the null distribution ($p < 0.001$).

\begin{figure}[H]
  \centering
  \safeincludegraphics[width=\linewidth]{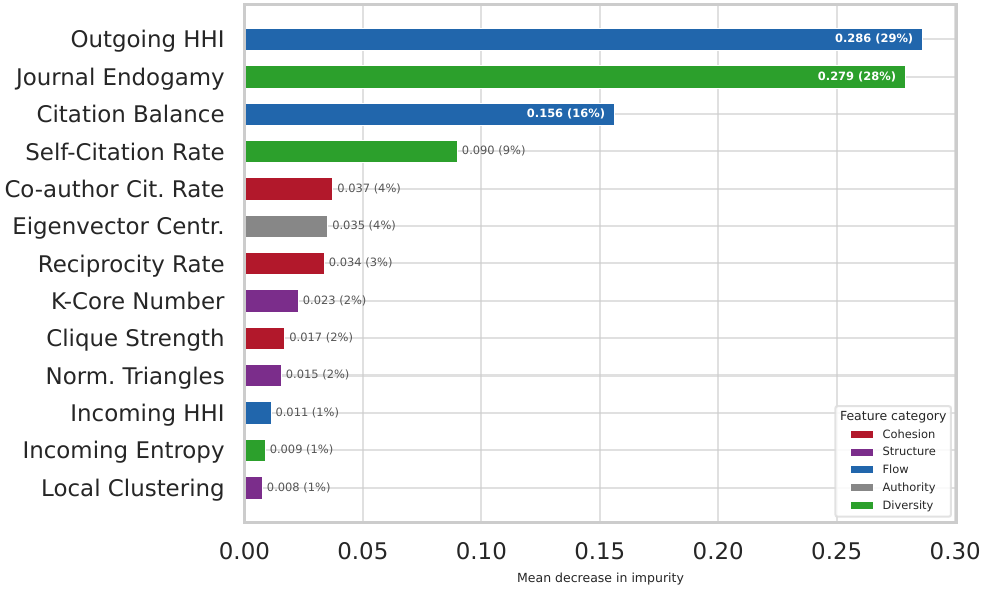}
  \caption{\textbf{Random Forest feature importances.}
    Horizontal bars show mean decrease in impurity for 13 of the
    14 behavioural features listed in Table~\ref{tab:features},
    sorted by importance.
    Bars are coloured by feature category (see legend).
    Inline labels give raw importance score and share of total.
  }
  \label{fig:feature_importance}
\end{figure}

In contrast to their high internal cohesion, Case authors exhibit significantly lower authority metrics. Controls exhibit \textbf{13$\times$ higher eigenvector centrality} than Cases ($p=3.4 \times 10^{-132}$), and \textbf{2.2$\times$ higher Citation Entropy}, indicating that while Cases have a similar ciation volume, they lack broad authoritative influence.
Note that the example pair in Table~\ref{tab:metrics-overview} both show near-zero eigenvector centrality at the $\times 10^{5}$ scale, which is consistent with the right-skewed distribution of this metric: the aggregate mean is driven by a small number of highly connected authors, and individual pairs can deviate substantially from the population average.
This divergence is further supported by the random forest feature importances in Figure~\ref{fig:feature_importance}, where cohesion and concentration metrics rank as the most discriminative features for tier classification, while authority metrics such as eigenvector centrality play a secondary role.

The Case--Control gap is explained mainly by distinct behavioral signatures. These patterns persist under conservative matched-pair comparisons: \textbf{12 out of 14} tested behavioral metrics remain significant after Benjamini-Hochberg FDR correction ($\alpha=0.05$) \parencite{Benjamini1995}.
This systemic disparity is not static: cumulative citation trajectories (2020--present) show the Case--Control gap \emph{widening} over time, indicating an accelerating divergence.

\subsection{Subject-Stratified Effects}
\label{sec:by-subject}

\begin{figure}[H]
  \centering
  \safeincludegraphics[width=0.95\linewidth]{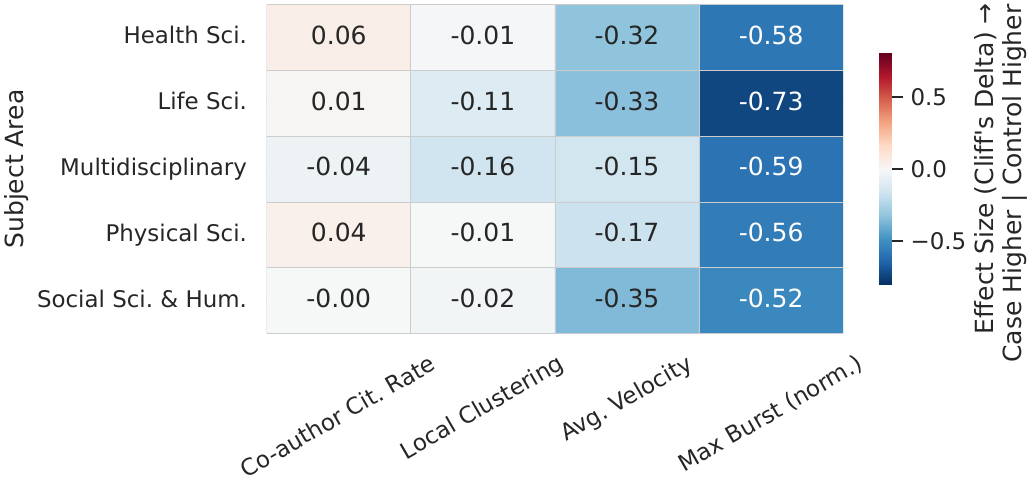}
  \caption{\textbf{Subject-Specific Effect Sizes (Cliff's $\delta$).}
    Each cell reports the effect size for a given metric within a subject area.
    Positive values (white--pale red) indicate that Cases exhibit higher values than Controls.
    Negative values (blue) indicate that Controls exhibit higher values than Cases.
    Co-author Citation Rate shows near-zero or slightly positive
    values across all subjects, while velocity and burstiness
    are uniformly negative, with Life Sciences displaying the
    largest deficit in burstiness ($\delta = -0.73$).
    \textit{Note.} While `Cohesion' (Co-author Rate) is universally high, other traits like `Burstiness' vary by field, confirming the `Subject--Specific Accents' hypothesis.}
  \label{fig:heatmap}
\end{figure}

Enhanced Case cohesion is a recurrent feature across subjects, but its expression varies by field (Figure~\ref{fig:heatmap}). In the health and life sciences, the divergence appears mainly as consistently higher co-author citation and stronger local clustering. In multidisciplinary journals and physical sciences, dense networking more often coincides with short, episodic activity waves. In the social sciences and humanities, cohesion signals are subtler, but sharper temporal concentrations of citations within short windows are more prominent.

Across fields, one-sided asymmetry measures remain generally subdued. Where differences emerge, they align more with transient speed-ups (velocity/bursts) than with persistent directional imbalance.

The signature of Case authors is greater network cohesion. Whether this manifests as intensified co-author referencing, increased local triadic closure, or episodic waves is field-dependent.

\subsection{Behavioral Archetypes}
K-means clustering on the standardised behavioural features reveals a clear three-way typology: \textbf{Central} (structurally prominent, well-embedded authors), \textbf{Independent} (typical, low-cohesion), and \textbf{Solo} (extreme self- and co-author citation).
Both tiers are dominated by the Independent profile, but the two peripheral clusters split along tier lines: Controls are nearly twice as likely to fall in the Central cluster (\num{19.0}\,\% vs \num{10.7}\,\%), whereas the Solo cluster is tiny but almost exclusive to Cases (27 of 29 authors). Thus K-means isolates a small, extreme Case-only archetype, while the broader cohesion gap reported in Section~\ref{sec:overall} manifests as a within-Independent shift rather than a distinct cluster.

\begin{figure}[H]
  \centering
  \safeincludegraphics[width=0.7\linewidth]{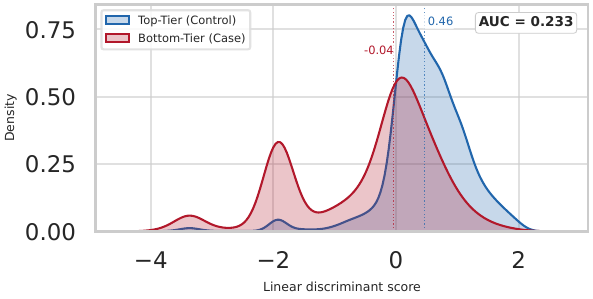}
  \caption{\textbf{Tier separability: LDA projection.}
    Kernel density estimates of the scores produced by projecting
    each author's standardised feature vector onto the single
    Linear Discriminant Analysis (LDA) axis that maximally separates
    {\color{caseRed}Case} from {\color{ctrlBlue}Control} authors.
    Dotted vertical lines mark within-tier medians.
    Negative scores correspond to the cohesion-driven regime
    (higher co-author citation, clustering, and reciprocity),
    while positive scores correspond to the independent regime
    (lower cohesion, typical citing behaviour).}
  \label{fig:lda}
\end{figure}

This divergence is consistent with clique-like behavior: Case authors follow multiple higher-cohesion strategies, whereas Controls converge on a single low-cohesion, outward-looking profile.
Figure~\ref{fig:lda} illustrates this separation: the Linear Discriminant Analysis (LDA) projection reveals minimal overlap between tiers, with Cases clustering toward the Cohesion-driven extreme and Controls toward the Authority-driven regime.

Three behavioral profiles emerge from the clustering:
\begin{enumerate}
  \item \textbf{Central} ($n = \num{2807}$; \num{10.7}\,\% of Cases, \num{19.0}\,\% of Controls):
        high structural prominence (k-core, triadic closure) combined with pronounced citation asymmetry;
        more prevalent among Controls, consistent with the embedded network position of top-tier authors.
  \item \textbf{Independent} ($n = \num{16026}$; \num{89.0}\,\% of Cases, \num{80.9}\,\% of Controls):
        low cohesion and connectivity; the dominant profile for both tiers.
  \item \textbf{Solo} ($n = 29$; \num{0.3}\,\% of Cases, ${<}\,\num{0.1}$\,\% of Controls):
        extreme self-citation, co-author citation, and reciprocity;
        nearly exclusive to Case authors (27 of 29). The cluster is small in absolute terms, so this archetype is descriptive rather than inferential.
\end{enumerate}
The Solo pattern—characterised by co-author citation rates nearly 200 times the population mean—appears almost exclusively among Cases.
Controls are proportionally more represented in the Central cluster, consistent with the structural prominence of top-tier authors.
Both tiers are dominated by the Independent baseline profile.

\subsection{Outlier Detection Results}\label{sec:outlier-results}

Using the hybrid detection strategy at the $4\sigma$ threshold, we identified \emph{277 high-confidence outliers}.
Of these, \textbf{93.5\%} are Case authors (\emph{Case purity}), confirming that the pipeline predominantly isolates authors whose structural profiles deviate substantially from the Control norm.

\begin{figure}[H]
  \centering
  \safeincludegraphics[width=0.6\linewidth]{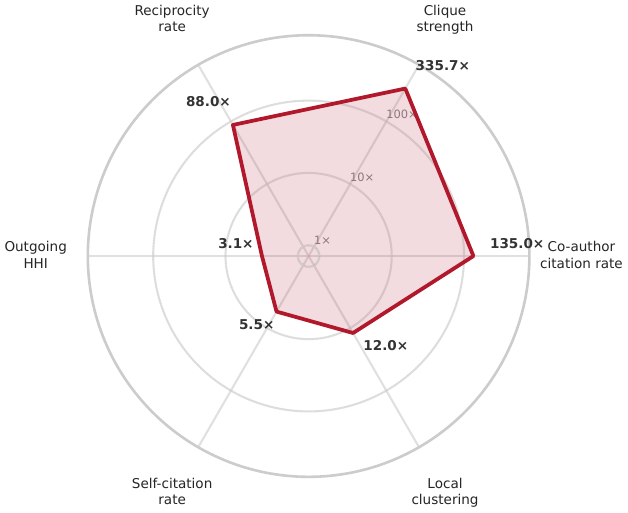}
  \caption{\textbf{Outlier behavioural fingerprint.}
    Radar chart of outlier fold-change ratios (log-scaled radial axis)
    relative to non-outlier authors across six citation-behaviour metrics.
    Values on each spoke give the outlier-to-normal mean ratio;
    concentric rings mark $1{\times}$, $10{\times}$, and $100{\times}$ baselines.
    The dashed inner ring corresponds to $1{\times}$ (no difference).}
  \label{fig:fingerprint}
\end{figure}

Relative to non-outlier authors, the detected syndicate members exhibit
many times higher cohesion signals (Fig.~\ref{fig:fingerprint}):
\textbf{Clique Strength $\approx 11\times$}, \textbf{Co-author Citation Rate $\approx 6.7\times$},
\textbf{Reciprocity $\approx 4.7\times$}, and \textbf{Tunnel Vision (Outgoing HHI) $\approx 3.1\times$}.
Note that the aggregate Case--Control fold-changes reported in Section~\ref{sec:overall} measure the same features across \emph{all} matched pairs, whereas the fingerprint ratios here compare the 277~flagged outliers to the remaining population.
These specific deviations constitute the unique \emph{structural fingerprint} of the flagged cohesion outliers.

\begin{table}[tb]
  \centering
  \caption{Sensitivity Analysis: Comparison of Detection Methods at $4\sigma$}
  \label{tab:contamination}
  \begin{threeparttable}
    \small
    \setlength{\tabcolsep}{6pt}
    \renewcommand{\arraystretch}{1.2}

    \begin{tabularx}{\linewidth}{@{} l c c c c @{}}
      \toprule
      \textbf{Method}                      & \textbf{Threshold} & \textbf{Outliers} & \textbf{Case Purity} & \textbf{Connected (\%)} \\
      \midrule
      Isolation Forest (Baseline)          & $4\sigma$          & 269               & 84\%                 & 69\%                    \\
      Cohesion Composite Only              & $4\sigma$          & 599               & 89\%                 & 26\%                    \\
      \textbf{Hybrid (IF $\cap$ Cohesion)} & \textbf{$4\sigma$} & \textbf{277}      & \textbf{94\%}        & \textbf{49\%}           \\
      \bottomrule
    \end{tabularx}
    \begin{tablenotes}
      \item \textit{Note:} The Hybrid method combines general multivariate anomalies (IF) with domain-specific cohesion scores, increasing Case Purity by +10 percentage points over the baseline.
    \end{tablenotes}
  \end{threeparttable}
\end{table}
\FloatBarrier

Table~\ref{tab:contamination} demonstrates the superiority of the hybrid approach.
While Isolation Forest alone achieves 84\% purity, augmenting it with the cohesion filter improves purity to 94\%.
This indicates that approximately 10\% of statistical outliers are legitimate prolific collaborators, who are successfully filtered out by the cohesion requirement.

\subsection{Network Forensics}

\begin{figure}[H]
  \centering
  \safeincludegraphics[width=\linewidth]{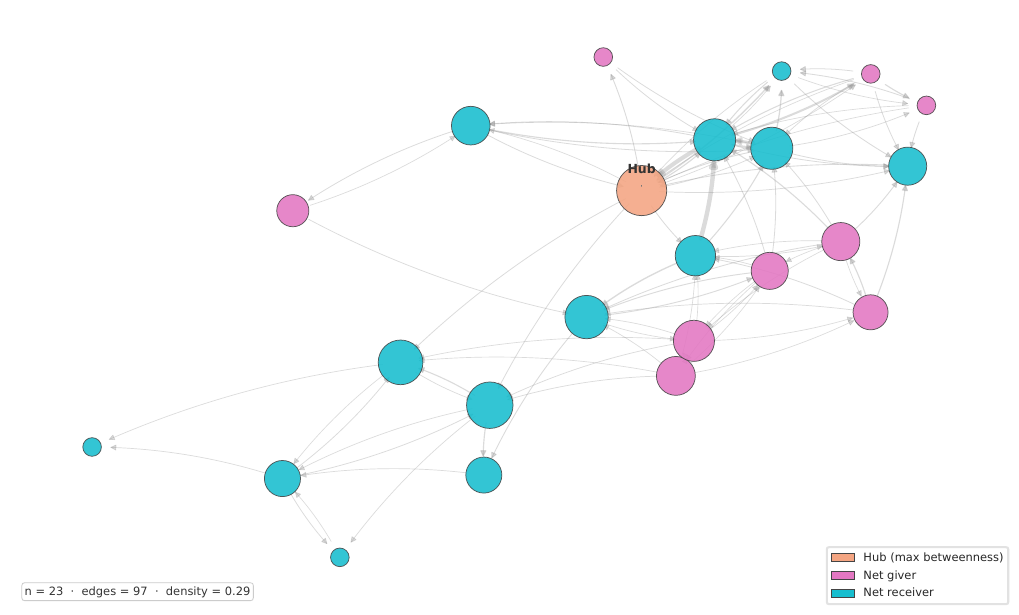}
  \caption{\textbf{Largest outlier citation syndicate ($n = 23$).}
  Force-directed layout of the internal directed citation network.
  Node size is proportional to betweenness centrality;
  \colorbox{hubSalmon!40}{\strut salmon} = hub (highest betweenness);
  \colorbox{giverPink!40}{\strut pink} = net giver (out-degree > in-degree);
  \colorbox{recvrCyan!40}{\strut cyan} = net receiver.
  Arrows indicate the direction of citation flow;
  edge width scales with the number of citations between two authors.
  The inset box reports node count, directed edge count, and
  graph density.}
  \label{fig:largest-syndicate}
\end{figure}

The largest outlier syndicate ($n=23$) exhibits a distinct hub-and-spoke topology rather than a dense, uniform mesh (Figure~\ref{fig:largest-syndicate}).
Within this group, the flow imbalance analysis reveals a clear hierarchy: a single \textbf{``Primary Beneficiary''} occupies the central position, receiving directed citation flows from peripheral \textbf{``Sycophants''} who define themselves by giving significantly more citations than they receive.

Temporal activity supports the orchestration hypothesis, showing a sudden-burst pattern with high internal citation intensity.
As shown in the timeline analysis, citation accumulation is not organic but driven by coordinated waves, particularly evident in 2021 and 2024.

The hub-and-spoke topology and coordinated temporal bursts reveal a pervasive ``Sycophant-Beneficiary'' flow imbalance that distinguishes Cases from Controls.

\subsection{Case--Control Segregation}

\begin{figure}[H]
  \centering
  \safeincludegraphics[width=0.85\linewidth]{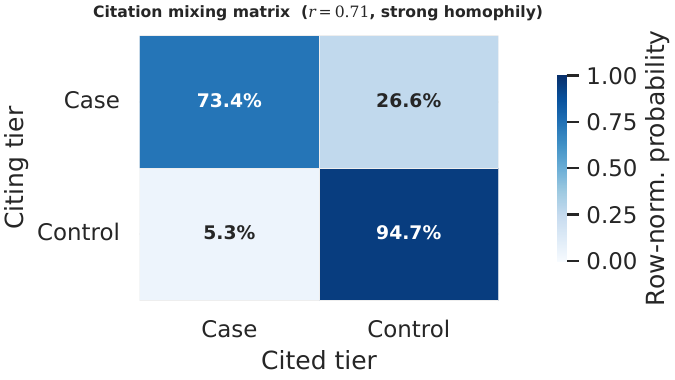}
  \caption{\textbf{Citation mixing matrix ($r = 0.71$, $Q = 0.97$; strong homophily).}
    Row-normalised proportion of citations directed within and
    across tiers.
    Cell values show the conditional probability that an author
    in the row tier cites an author in the column tier.
    The Brewer RdBu colormap is centred at 0.5.
    Below: diagonal average (84.0\%) summarises overall within-tier
    preference.}
  \label{fig:mixing_matrix}
\end{figure}

The distinction between low-- and high-impact venues extends beyond simple density; it reflects extreme \emph{segregation}.
The high assortativity coefficient ($r=0.71$) \parencite{Newman2002} and modularity ($Q=0.97$) \parencite{NewmanGirvan2004} confirm a \textbf{``Two Worlds'' hypothesis}, where Case authors operate in closed citation economies with minimal external engagement (Figure~\ref{fig:mixing_matrix}).
Flagged outliers concentrate at the high-asymmetry, high-burst corner of the anomaly landscape, a region essentially unpopulated by Controls.
In this sense, citation cliques are not merely organic clusters but isolated silos disconnected from the broader scientific discourse.

\subsection{Cross-referencing Case Authors}\label{sec:suspicious-authors}

To move from aggregate outlier statistics to actionable intelligence,
we cross-reference the 277 hybrid outliers against real-world author
identities and publication records.
Of these, 189 (68.2\%) had publicly resolvable ORCID profiles
and are analysed below; the remaining 88 lacked public ORCID records
or returned incomplete metadata.
Each ORCID identifier is resolved to a name via the ORCID public
API,\footnote{\url{https://pub.orcid.org}} and publication portfolios
are reconstructed from \texttt{impact.db} by joining
\texttt{work\_authors} to \texttt{works} and resolving journal titles
through the \texttt{journal\_names} catalogue.

We define a composite \emph{outlier score} as the weighted sum
of population-normalised $z$-scores across six behavioural features:
\begin{equation}\label{eq:susp}
  S_i = \sum_{k=1}^{6} w_k \cdot z_{ik},
  \qquad
  z_{ik} = \frac{x_{ik} - \bar{x}_k}{\sigma_k},
\end{equation}
where $x_{ik}$ is the raw value of feature~$k$ for author~$i$,
$\bar{x}_k$ and $\sigma_k$ are the population mean and standard
deviation, and the weights $w_k$ rank the six features by their
discriminative strength between Case and Control authors.
To avoid overfitting the weights to the sample, we did not optimise
them numerically; instead, each feature was assigned an integer or
half-integer weight on the grid $\{2.0, 3.0, 3.5, 4.0\}$ in proportion to
two complementary evidence sources:
(i)~Random-Forest mean decrease in impurity for Case--Control
classification (Figure~\ref{fig:feature_importance}), and
(ii)~the Wilcoxon rank-biserial correlation from the matched-pair
tests.
Features ranking top on \emph{both} criteria received the largest
weight; features ranking high on one criterion only, or significant
with a smaller effect size, received a lower weight.
The resulting schedule is:
co-author citation rate ($w=4.0$; largest effect size and top RF importance),
clique strength ($3.5$) and reciprocity rate ($3.5$; large effects, second-tier RF importance),
outgoing HHI ($3.0$; concentration signal complementary to the cohesion metrics),
self-citation rate ($2.0$) and journal endogamy rate ($2.0$;
smaller but still significant Case--Control differences).
The score is a heuristic ranking device rather than a fitted model;
the final outlier flag is issued only when $S_i$ and the independent
Isolation Forest both exceed their thresholds.

Authors are ranked by $S_i$ in descending order.
An individual feature is flagged when its $z$-score exceeds $3\sigma$
or $5\sigma$.

\textbf{Key Findings.}
Of the 189 outliers, 162 (85.7\%) belong to the Case
group, confirming the precision of the Isolation Forest detector.
Outliers concentrate in Health Sciences ($n=77$) and Physical Sciences ($n=71$),
followed by Social Sciences ($n=46$) and Life Sciences ($n=3$);
per-subject counts sum to 197 because eight authors are flagged in
two subjects each.
A total of 131 outliers (69.3\%) belong to at least one syndicate
(connected component among mutual outlier citation links), with the
largest syndicate comprising 23~members (see Figure~\ref{fig:largest-syndicate}).

Table~\ref{tab:suspicious-top10} presents the ten highest-scoring
authors.
All ten exhibit co-author citation rates exceeding $5\sigma$;
seven also show reciprocity rates beyond $5\sigma$.
The two top-ranked authors --- both in agricultural engineering---
share an identical outlier score ($S = 658.6$) and belong to
the same 23-member syndicate, consistent with coordinated behaviour.
The presence of authors writing in Korean, Russian, Ukrainian, and
Portuguese underscores the international scope of the phenomenon.

Table~\ref{tab:author-audit} provides a publication-level audit of
the same ten authors.
Typical portfolios are small (4--14 works, median~5), concentrated
in a single regional journal, and characterised by very few distinct
citation partners (1--5) yet elevated reciprocal citation links,
a pattern consistent with closed-loop mutual citation within a
small community of co-authors.

\begin{table*}[t]
  \centering
  \caption{Top-10 Case authors ranked by composite
    outlier score $S$ (Eq.~\ref{eq:susp}).}
  \label{tab:suspicious-top10}
  \scriptsize
  \begin{tabularx}{\linewidth}{rl l r r X}
    \toprule
    Rank & ORCID                        & Subject & $S$   & Works & Primary flags                                                 \\
    \midrule
    1    & \texttt{\dots5580} & Eng.    & 658.6 & 12    & co-auth.\ cit., clique str., recip.\ ($>5\sigma$)             \\
    2    & \texttt{\dots6290} & Eng.    & 658.6 & 13    & co-auth.\ cit., clique str., recip.\ ($>5\sigma$)             \\
    3    & \texttt{\dots3400} & Med.    & 361.7 & 4     & co-auth.\ cit., recip., out.\ HHI ($>5\sigma$)                \\
    4    & \texttt{\dots5180} & Soc.    & 207.6 & 14    & co-auth.\ cit.\ ($>5\sigma$); self-cit.\ ($>3\sigma$)         \\
    5    & \texttt{\dots9510} & Eng.    & 204.9 & 4     & co-auth.\ cit., clique str., recip.\ ($>5\sigma$)             \\
    6    & \texttt{\dots3020} & Eng.    & 203.4 & 5     & co-auth.\ cit., clique str., recip.\ ($>5\sigma$)             \\
    7    & \texttt{\dots4470} & Eng.    & 203.0 & 5     & co-auth.\ cit., recip., out.\ HHI ($>5\sigma$)                \\
    8    & \texttt{\dots8870} & Eng.    & 184.7 & 4     & co-auth.\ cit., recip.\ ($>5\sigma$); endog.\ ($>5\sigma$)    \\
    9    & \texttt{\dots9530} & Eng.    & 180.4 & 5     & co-auth.\ cit., recip.\ ($>5\sigma$); self-cit.\ ($>3\sigma$) \\
    10   & \texttt{\dots6810} & Eng.    & 180.4 & 4     & co-auth.\ cit., recip.\ ($>5\sigma$); self-cit.\ ($>3\sigma$) \\
    \bottomrule
  \end{tabularx}

  \medskip
  {\footnotesize\textit{Note.}
  Flags indicate features exceeding $3\sigma$ or $5\sigma$
  of the population mean.
  These patterns are \emph{statistically anomalous}; they do not
  constitute proof of misconduct.
  Full author profiles are accessible via \texttt{https://orcid.org/[ORCID]}.}
\end{table*}

\begin{table*}[t]
  \centering
  \caption{Publication audit profiles for top-10 outlier Case authors.}
  \label{tab:author-audit}
  \scriptsize
  \begin{tabularx}{\linewidth}{rl rrr rrX}
    \toprule
    Rank & ORCID                        & Works & Years      & Out & In & Recip. & Primary journal                  \\
    \midrule
    1    & \texttt{\dots5580} & 12    & 2020--2024 & 1   & 3  & 1      & Bull.\ Sumy Nat.\ Agrar.\ Univ.\ (6)  \\
    2    & \texttt{\dots6290} & 13    & 2020--2024 & 1   & 3  & 1      & Anim.\ Breed.\ Genet.\ (7)            \\
    3    & \texttt{\dots3400} & 4     & 2020--2023 & 1   & 1  & 1      & Laser Medicine (2)                     \\
    4    & \texttt{\dots5180} & 14    & 2020--2024 & 2   & 0  & 0      & Exacta (5)                             \\
    5    & \texttt{\dots9510} & 4     & 2020--2022 & 3   & 5  & 1      & Korean J.\ Soil Sci.\ Fert.\ (4)      \\
    6    & \texttt{\dots3020} & 5     & 2020--2022 & 3   & 5  & 1      & Korean J.\ Soil Sci.\ Fert.\ (5)      \\
    7    & \texttt{\dots4470} & 5     & 2020--2022 & 1   & 1  & 1      & Acta Sci.\ Agronomy (2)                \\
    8    & \texttt{\dots8870} & 4     & 2021--2023 & 2   & 1  & 1      & Korean J.\ Plant Taxon.\ (3)           \\
    9    & \texttt{\dots9530} & 5     & 2021--2023 & 2   & 1  & 1      & Sci.\ Notes Kazan Vet.\ Acad.\ (2)    \\
    10   & \texttt{\dots6810} & 4     & 2022--2023 & 2   & 1  & 1      & Int.\ J.\ Vet.\ Med.\ (2)             \\
    \bottomrule
  \end{tabularx}

  \medskip
  {\footnotesize\textit{Note.}
    Out = distinct sample authors with ORCIDs cited by this author within 2020--2024 (self-citations excluded);
    In = distinct sample authors with ORCIDs who cite this author;
    Recip.\ = reciprocal citation partners.
    Counts are restricted to the matched-pair sub-network and to ORCID-resolved citation links, so they substantially undercount each author's full reference list.
    ``Primary journal'' is the journal in which the author published
    the most works within the 2020--2024 study window.}
\end{table*}

The concentration of outlier behaviour in small, regional journals
many of which do not appear in major indexing databases, suggests that coordinated citation inflation is disproportionately
prevalent in venues with limited editorial oversight.
The publication-audit data show that the highest-scoring authors
maintain very small portfolios (median~5 works), yet achieve
inflated citation metrics through closed-loop mutual referencing.
This pattern is consistent with prior work on ``citation stacking''
\parencite{heneberg2016} and suggests that journal-level surveillance
of incoming self-citation rates from small author cliques could serve
as an effective early-warning signal.

While broad asymmetry metrics are generally subdued, the internal structure of detected syndicates reveals a distinct \textbf{flow imbalance}.
As seen in the largest syndicate ($n=23$), a hub-and-spoke topology emerges where peripheral members (``Sycophants'') act as net givers to central ``Beneficiaries''.
This suggests that the observed cohesion serves a functional purpose: the coordinated funneling of credit to central figures, rather than reciprocal exchange among equals.

These findings extend the concept of \emph{citation orchestration} \parencite{Evdaimon2024}, showing that orchestration manifests as persistent reciprocity and dense closure.
The limited cross-tier mixing supports concerns about \emph{home bias} \parencite{Qiu2025} and institutional siloing.
However, the specificity of our \emph{hybrid detection} (93.5\% purity) suggests that these patterns in low-impact venues are structural anomalies distinct from legitimate topic specialization.

The results highlight that aggregate impact metrics hide structural manipulation.
Evaluation systems must incorporate \textbf{cohesion-aware metrics} --- specifically \emph{co-author citation rates} and \emph{flow imbalance}--- to distinguish organic collaboration from dense internal referencing.
The success of the hybrid pipeline suggests that combining multivariate anomaly detection (Isolation Forest) with targeted cohesion signals can effectively flag insular publication ecosystems.

Overall, the dominance of segregation and flow imbalance suggests that low-impact venues foster \textbf{closed economies} of citation.
Understanding these patterns is essential for designing fair bibliometric indicators that reward genuine impact rather than inward-looking visibility loops.

\section{Limitations and Future Work}
\label{sec:limitations}

This study relies on Crossref metadata (2020--2024), which vary in completeness.
While the hybrid filter significantly reduces false positives, the distinction between a ``cartel'' and a hyper-specialized research group remains a probabilistic assessment.
The LLM-based journal classification (Appendix~\ref{app:llm}) was not validated against an independent ground truth; misclassification of some journals could shift the boundary between Case and Control authors.
Additionally, journal classification relied solely on a proprietary
model (\texttt{gpt-4.1-2025-04-14}) without an open-weight baseline
comparison and without independent human validation of the resulting
labels \parencite{LLMGuidelines2024}.
Moreover, citation links are analyzed structurally, without text-level context to infer the motivation of citing \parencite{Liu2024}.

\paragraph{Ethical considerations.}
Where individual author profiles are discussed (Tables~\ref{tab:suspicious-top10} and~\ref{tab:author-audit}), the underlying data consist exclusively of publicly available Crossref metadata and self-declared ORCID identifiers.
Identifying anomalous citation patterns in publicly funded research serves a legitimate accountability function, consistent with established practice in the retraction and research-integrity literature \parencite{Biagioli2020,Besancon2024}.
Although explicit author names and institutional affiliations are omitted from the text and tables for discretion, the underlying publication records remain fully transparent and verifiable through the provided ORCID identifiers.

Future work could combine the structural fingerprint with author disambiguation and profile-level inspection to move from structural suspicion to targeted auditing.
Extending the temporal coverage could clarify whether the observed ``Two Worlds'' segregation is widening over time.
Experimental audits in selected venues could also test whether editorial interventions can dismantle the echo chambers identified in this analysis.

\section{Summary and Conclusions}
\label{sec:conclusion}

We developed a subject-aware hybrid pipeline that integrates LLM-based journal classification, per-subject Eigenfactor scores, and a composite outlier detection strategy.
This design allowed us to isolate behavioral signatures with author productivity (by subject-specific $h_5$) held constant.

The evidence points to a highly structured phenomenon.
In the aggregate Case--Control comparison, authors publishing predominantly in low-impact journals cite co-authors \textbf{6.7$\times$ more frequently}; among the 277~flagged outliers the lift is even starker, with \textbf{11$\times$ greater clique strength} relative to non-outliers.
Rather than random noise, outlier syndicates display a \textbf{hub-and-spoke topology}, where peripheral ``Sycophants'' funnel citations to central ``Beneficiaries'' through coordinated temporal citation bursts.
The flow imbalance analysis confirms that Case authors systematically give more citations than they receive.
Cumulative citation trajectories over the 2020--2024 window suggest that the gap between tiers may be widening, although a longer observation period is needed to confirm this trend.

Controlling for subject and productivity, the hybrid detection strategy identifies structurally anomalous citation groups with 93.5\% Case purity (proportion of flagged authors belonging to the Case group).
These groups are characterized not by simple asymmetry, but by a specific cohesion fingerprint: extreme co-author citation rates, dense local clustering, and tunnel vision.
Consequently, some low-impact venues appear to foster \textbf{segregated, inward-looking citation economies} whose structural properties are associated with distortion of bibliometric indicators.

\section{Author Contributions}
Conceptualization: P.-A.S., D.S.; Methodology: P.-A.S.; Software: P.-A.S., G.A.;
Data curation: P.-A.S., G.A.; Validation: P.-A.S., G.A., D.S.; Formal analysis: P.-A.S., G.A.;
Visualization: P.-A.S.; Writing -- original draft: G.A.; Writing -- review \& editing: P.-A.S., G.A., D.S.;
Supervision: D.S.; Project administration: D.S.

\section{Competing Interests}
The authors declare no conflict of interest.

\section{Funding Information}
This research did not receive a specific grant from any funding agency in the public, commercial or non-profit sectors.

\section{Data and Code Availability}
\label{sec:data}
All essential data and code for this study are openly available at Zenodo under a CC~BY license \parencite{dataset_zenodo}.
The analysis is implemented in a single, self-contained script (\texttt{citation\_analysis.py}) that reproduces all results in one invocation: it loads the pre-built \texttt{rolap.db} and \texttt{impact.db} databases, runs the statistical comparisons (Section~\ref{sec:overall}), generates all eight publication figures, and produces the author outlier tables (Section~\ref{sec:suspicious-authors}).
The citation databases and the derivative tables can be reproduced
through the 2024 Crossref public data set by means of the supplied
Unix Makefile and SQL scripts.

\printbibliography

\appendix

\section{Feature Definitions}
\label{app:features}

Table~\ref{tab:features} provides formal definitions for all behavioural
metrics used in the analysis.  In the formulas below, $w_{ij}$ denotes the
fractional citation weight from author~$i$ to author~$j$
(see Appendix~\ref{app:params}), and sums run over non-self edges
unless stated otherwise.

  {\singlespacing\small
    \begin{longtable}{@{}l p{0.25\textwidth} >{\RaggedRight\arraybackslash}p{0.38\textwidth}@{}}
      \caption{\textbf{Behavioural feature definitions.}
        All metrics are computed per author over the 2020--2024 citation network.
        $N_{\text{out}}$~= number of distinct cited authors;
        $N_{\text{in}}$~= number of distinct citing authors;
        $p_j = w_{ij} / \sum_k w_{ik}$ (outgoing share);
        $q_j = w_{ji} / \sum_k w_{ki}$ (incoming share).}
      \label{tab:features}                                                                                                                                                                                                                                                 \\
      \toprule
      \textbf{Feature} & \textbf{Formula}                                                                                                                                                                                                        & \textbf{Interpretation} \\
      \midrule
      \endfirsthead
      \multicolumn{3}{l}{\small\textit{Table~\ref*{tab:features} continued}}                                                                                                                                                                                               \\
      \toprule
      \textbf{Feature} & \textbf{Formula}                                                                                                                                                                                                        & \textbf{Interpretation} \\
      \midrule
      \endhead
      \midrule
      \multicolumn{3}{r}{\small\textit{Continued on next page}}                                                                                                                                                                                                            \\
      \endfoot
      \bottomrule
      \endlastfoot
      Self-citation rate
                       & $\displaystyle\frac{\sum w_{ii}}{\sum_j w_{ij}}$
                       & Share of outgoing citation weight directed at the author's own works \parencite{Aksnes2003}.                                                                                                                                                     \\
      \addlinespace
      Co-author citation rate
                       & $\displaystyle\frac{\sum_{j\in C_i} w_{ij}}{\sum_{j\neq i} w_{ij}}$
                       & Among non-self citations, proportion citing former co-authors ($C_i$).                                                                                                                                                                            \\
      \addlinespace
      Citation balance
                       & $\displaystyle\frac{w_{\text{out}} - w_{\text{in}}}{w_{\text{out}} + w_{\text{in}} + \epsilon}$
                       & Net giver/receiver bias; positive~$\Rightarrow$ net giver. Bounded $(-1,+1)$, comparable across activity levels.                                                                                                                                  \\
      \addlinespace
      Eigenvector centrality
                       & $\lambda_1$-eigenvector of $G$
                       & Global prestige score: leading eigenvector of the weighted directed citation graph \parencite{Bonacich1987}.                                                                                                                                       \\
      \addlinespace
      Journal endogamy
                       & $\displaystyle\frac{|\{\text{refs with same ISSN}\}|}{|\{\text{all refs}\}|}$
                       & Proportion of references citing papers in the same journal.                                                                                                                                                                                       \\
      \addlinespace
      Citation entropy
                       & $H = -\sum_j q_j\ln q_j$
                       & Shannon diversity \parencite{Shannon1948} of incoming citation sources. Low~$\Rightarrow$ concentrated.                                                                                                                                          \\
      \addlinespace
      Citation HHI (in)
                       & $\sum_j q_j^{\,2}$
                       & Herfindahl--Hirschman concentration index \parencite{Hirschman1964} of incoming citations.                                                                                                                                                       \\
      \addlinespace
      Reciprocity rate
                       & $\displaystyle\frac{|\{j : w_{ij}>0 \wedge w_{ji}>0\}|}{|\{j : w_{ij}>0\}|}$
                       & Share of cited peers who also cite back \parencite{Newman2002}.                                                                                                                                                                                  \\
      \addlinespace
      Outgoing HHI
                       & $\sum_j p_j^{\,2}$
                       & Herfindahl--Hirschman concentration \parencite{Hirschman1964} of outgoing citations (``tunnel vision'').                                                                                                                                         \\
      \addlinespace
      Clustering coeff.
                       & \resizebox{\linewidth}{!}{$\displaystyle\frac{2\,|\{(j,k)\colon j,k\in\mathcal{N}_i,\,(j,k)\in E\}|}{d_i(d_i-1)}$}
                       & Fraction of an author's neighbours that are themselves connected \parencite{WattsStrogatz1998}. Computed on the undirected projection of the internal citation graph.                                                                             \\
      \addlinespace
      Triangles (norm.)
                       & $\displaystyle\frac{\Delta_i}{\sum_j w_{ij} + \sum_j w_{ji} + 1}$
                       & Triangle count normalised by total citation activity.                                                                                                                                                                                             \\
      \addlinespace
      $k$-core number
                       & $\max\{k : i \in H_k\}$
                       & Largest $k$ such that author~$i$ belongs to the $k$-core of the undirected citation graph \parencite{Seidman1983}.                                                                                                                               \\
      \addlinespace
      Clique strength
                       & clustering $\times$ co-author citation rate
                       & Joint cohesion signal: dense neighbourhood \emph{and} preferential co-author citing.                                                                                                                                                              \\
      \addlinespace
      Norm.\ burst intensity\textsuperscript{\dag}
                       & $\displaystyle\frac{\max_j w_{ij}}{\text{in\_strength} + 1}$
                       & Largest single-peer citation weight normalised by total incoming volume; inspired by burst-detection methods for streaming events \parencite{Kleinberg2003}. Available for \num{7711} of \num{9431} pairs; excluded from the Random Forest owing to 18\% missing data (see Section~\ref{sec:outlier-proc}).                           \\
    \end{longtable}
  }

\section{Matching Procedure and Detection Parameters}
\label{app:params}

\subsection{Author Classification}\label{sec:author-classification}
Authors were classified based on their publication portfolio:
\begin{itemize}
  \item \textbf{Case}: $\geq70\%$ of papers published in journals below the 25th subject-specific Eigenfactor percentile, with a minimum of 3~papers.
  \item \textbf{Control}: $\geq70\%$ of papers published in journals above the 75th subject-specific Eigenfactor percentile, with a minimum of 3~papers.
  \item \textbf{Other}: authors not meeting either threshold.
\end{itemize}
The 70\% threshold ensures that classified authors are predominantly active
in one venue stratum while tolerating occasional cross-tier publications
(e.g.\ a single paper in a high-impact journal by an otherwise low-tier
author).  Lower thresholds (50--60\%) admit too many mixed-portfolio
authors, diluting the behavioural contrast; higher thresholds ($\geq$80\%)
reduce sample sizes substantially and bias the cohort toward single-journal
authors.  The 70\% cut-off balances purity against statistical power and
is consistent with the majority-venue classification used in large-scale
author-level analyses \parencite{Ioannidis2019,Ioannidis2024extreme}.

\subsection{Matched-Pair Construction}
Case--Control pairs were formed via greedy one-pass matching within each subject area.
For each Case author the algorithm selected the closest available Control author by subject-specific $h_5$~index.
Productivity was bucketed by integer division ($b = \lfloor h_5/3 \rfloor$); candidates with $|b_{\text{case}} - b_{\text{control}}| \leq 1$ (i.e., $\pm3$~$h_5$ points) were eligible.
Among eligible candidates the one minimising $|h_{5,\text{case}} - h_{5,\text{control}}|$ was selected.
Each (ORCID, subject) pair was used at most once (matching without replacement).
This procedure yielded \num{9431} matched pairs across five subject areas (Table~\ref{tab:sampling-funnel}).

\subsection{Fractional Citation Weights}
Each citation edge between a citing work with $n_c$~authors and a cited work with $n_a$~authors received weight
\[
  w = \frac{1}{n_c \times n_a}\,.
\]
This fractional counting scheme \parencite{PerianeRodriguez2016}
down-weights edges from heavily co-authored papers, preventing a
single multi-author article from dominating an individual's citation
profile.
Self-loops ($i = j$) were retained and tracked via a Boolean flag for the self-citation rate metric; all other metrics were computed on non-self edges.
The temporal window covers works published between 2020 and 2024 inclusive,
yielding \num{2525512} weighted citation edges (\num{2507806} non-self,
\num{17706} self-citations) among \num{18862} study authors.

\subsection{Outlier Detection}
Table~\ref{tab:if-params} summarises the Isolation Forest hyper-parameters.
The model was trained per-subject on 13 of the 14 scaled features listed in Table~\ref{tab:features}
(excluding normalised burst intensity; see Section~\ref{sec:outlier-proc}).

\begin{table}[htbp]
  \centering
  \captionsetup{justification=centering}
  \caption{\textbf{Isolation Forest hyper-parameters.}}
  \label{tab:if-params}
  \begin{tabular}{@{}ll@{}}
    \toprule
    \textbf{Parameter}     & \textbf{Value}                 \\
    \midrule
    \texttt{n\_estimators} & 200                            \\
    \texttt{contamination} & 0.01                           \\
    \texttt{max\_samples}  & \texttt{auto} ($\min(256, n)$) \\
    \texttt{max\_features} & 1.0 (all features)             \\
    \texttt{bootstrap}     & False                          \\
    \texttt{random\_state} & 42                             \\
    \bottomrule
  \end{tabular}
\end{table}

\paragraph{Cohesion composite score.}
A secondary filter computes a weighted $z$-score relative to the Control population:
\[
  S_i = \sum_{k=1}^{6} w_k \cdot z_{ik},
  \qquad
  z_{ik} = \frac{x_{ik} - \bar{x}_k}{\sigma_k}\,,
\]
with weights: co-author citation rate ($4.0$), clique strength ($3.5$), reciprocity rate ($3.5$),
outgoing HHI ($3.0$), self-citation rate ($2.0$), journal endogamy ($2.0$).
Standardisation uses the population mean and standard deviation.
An author was flagged only if both the Isolation Forest anomaly score exceeded its per-subject threshold \emph{and} $S_i > 4\sigma$.

\section{Data Derivation and Sampling}
\label{app:sampling}

The analytical dataset was derived through a multi-stage pipeline.
This appendix documents the full data flow and the filtering decisions at each stage.

\subsection{Stage~1: Raw Data Ingestion}
Crossref metadata were ingested via \textit{alexandria3k} \parencite{Spinellis2023}, retaining works published between 2020 and 2024.
The following fields were extracted: DOI, publication year, page information, print and electronic ISSN, reference lists (DOI-level), and author ORCID identifiers.

\subsection{Stage~2: Journal-Level Metrics}
Works were linked to journals through normalised ISSNs.
A citability filter removed non-article items (single-page entries such as editorials and news pieces).
For each journal, two-year and five-year citation means, the $h_5$~index, network centrality (Eigenfactor) \parencite{West2008}, prestige rank (SJR-like) \parencite{GonzalezPereira2010}, mean article score, and context-normalised impact were computed.
Journal citation networks were constructed from references in the reference year (2024) to publications in the preceding two-, three-, or five-year windows, depending on the metric.

\subsection{Stage~3: Journal Subject Classification}
Each journal was assigned to one of five broad subjects using an LLM-based classifier (see Appendix~\ref{app:llm}).
The final mapping covers approximately \num{15000} ISSNs.
Within each subject, the 25th and 75th Eigenfactor percentiles were computed to define \emph{low-impact} and \emph{high-impact} thresholds, respectively.

\subsection{Stage~4: Author Population and Sampling}
To keep computation tractable, the author population was sampled by retaining only ORCIDs whose last digit is~\texttt{0} (approximately 10\% of all ORCID-bearing authors in the Crossref snapshot).
This deterministic filter preserves the distribution of author characteristics because the ORCID check digit is algorithmically assigned and uncorrelated with research behaviour.

\subsection{Stage~5: Author Classification and Matching}
Within the sampled population, authors were classified per subject (detailed in Section~\ref{sec:author-classification}).
Authors not meeting either criterion were excluded.
A subject-specific $h_5$~index was computed for each remaining author.
Case authors were then matched to Control authors via greedy one-pass matching within subject, as described in Appendix~\ref{app:params}.

\subsection{Stage~6: Final Dataset}
The matching procedure yielded \num{9431} Case--Control pairs across five subjects (Table~\ref{tab:sampling-funnel}).
For the maximum citation burst metric, \num{7711} pairs were available due to data sparsity.
Directed author-to-author citation networks were constructed for matched authors using fractional edge weights (Appendix~\ref{app:params}).

\begin{table}[htbp]
  \centering
  \captionsetup{justification=centering}
  \caption{\textbf{Matched pairs by subject area.}}
  \label{tab:sampling-funnel}
  \begin{tabular}{@{}lr@{}}
    \toprule
    \textbf{Subject}              & \textbf{Pairs}      \\
    \midrule
    Health Sciences               & \num{3776}          \\
    Life Sciences                 & \num{2543}          \\
    Physical Sciences             & \num{2262}          \\
    Social Sciences \& Humanities & 518                 \\
    Multidisciplinary             & 332                 \\
    \midrule
    \textbf{Total}                & \textbf{\num{9431}} \\
    \bottomrule
  \end{tabular}
\end{table}

\section{LLM-Based Journal Classification}
\label{app:llm}

Journal subject assignment used OpenAI \texttt{gpt-4.1}
(snapshot \texttt{gpt-4.1-2025-04-14}, accessed June~2025)
with temperature~0 and deterministic seeding.

The prompt supplied the journal name and, where available, a representative abstract and article titles, together with few-shot exemplars for each category.
Decision rules prioritised abstract content over article titles and journal names to reduce ambiguity.
Predictions were cached deterministically to a JSON file; the final mapping covers approximately \num{15000} ISSNs stored in the \texttt{issn\_subjects} table of the study database.
No independent ground-truth validation against human annotators was
performed, and no open-weight model was used as a baseline;
both are acknowledged as limitations (Section~\ref{sec:limitations}).

\end{document}